\RequirePackage{fix-cm}

\documentclass[twocolumn,epjc3]{svjour3}  
\smartqed  % flush right qed marks, e.g. at end of proof
\RequirePackage{graphicx}
\usepackage{xcolor}
\usepackage{amssymb}
\usepackage{epsfig}
\usepackage{epstopdf}
\usepackage{graphicx}
\usepackage{amsmath}

\colorlet{darkgreen}{green!50!black}
\colorlet{brightyellow}{yellow!75!red}
\colorlet{orange}{red!50!yellow}
\colorlet{darkred}{red!80!black}
\colorlet{darkblue}{blue!50!black}

\usepackage{soul}

\journalname{Eur. Phys. J. C}
\begin{document}

\title{Euclidean to Minkowski Bethe-Salpeter amplitude and observables}%\thanksref{t1}



\author{J.~Carbonell\thanksref{e1,addr1}
        \and
        T.~Frederico\thanksref{e2,addr2} %etc.
        \and
        V.A.~Karmanov\thanksref{e3,addr3} %etc.        
}
\thankstext{e1}{e-mail: carbonell@ipno.in2p3.fr}
\thankstext{e2}{Corresponding author; e-mail: tobias@ita.br}
\thankstext{e3}{e-mail: karmanov@sci.lebedev.ru}
%\thankstext{e1}{e-mail: fauthor@example.com}


\institute{Institut de Physique Nucleaire, Universit\'e Paris-Sud, IN2P3-CNRS, 91406 Orsay Cedex, France \label{addr1}
           \and
Instituto Tecnol\'ogico de Aeron\'autica, DCTA, 12228-900, S. Jos\'e dos Campos,~Brazil \label{addr2}
           \and
Lebedev Physical Institute, Leninsky Prospekt 53, 119991 Moscow, Russia \label{addr3}
}
\date{Received: date / Accepted: date}

\maketitle
\begin{abstract}
We propose a method to reconstruct the  Bethe-Salpeter  amplitude  in Minkowski space
given the Euclidean Bethe-Salpeter amplitude  --  or alternatively  the Light-Front wave function --  as input.
The method is based  on the numerical  inversion of the Nakanishi integral representation and computing the corresponding weight function.
This inversion procedure is, in general, rather unstable, and we propose several ways  to  considerably reduce the instabilities. 
In terms of the Nakanishi weight function, one 
can easily compute the BS amplitude, the LF wave function and  the electromagnetic  form factor.  
The latter ones are very stable  in spite of residual instabilities in the weight function. 
This procedure allows both, to continue the Euclidean BS solution  in the Minkowski space and to obtain a BS amplitude from  a LF wave function.
\keywords{Bethe-Salpeter equation, Nakanishi Representation, Light-Front}
\end{abstract}

%%%%%%%%%%%%%%%%%%%%%%%%%%
\section{Introduction}

Among the methods in quantum field theory and quantum mechanical approaches to relativistic few-body systems ,
the Bethe-Salpeter (BS) equation \cite{SB_PR84_51}, based on  
first principles and existing already for more than sixty years, remains  rather popular. 

The solution of the BS equation in Euclidean space is much more simple than in Minkowski one. 
Rather often the Euclidean solution is  enough, mainly when one is interested in finding the binding energy. 
However, in many cases ({\it e.g.} for calculating electromagnetic  (EM)  form factors) one needs the Minkowski space amplitude
or, equivalently, the Euclidean one with complex arguments \cite{Guernot_Eichmann}. 
During the recent years,  new methods to find the Minkowski solution for the two-body bound state
were developed and proved their efficiency, at least for  simple 
kernels like the one-boson exchange  (OBE). 
Some of them \cite{KusPRD,bs1,FrePRD12,fsv_2014} are based on the Nakanishi 
integral representation \cite{nakanishi} of the BS amplitude and provide,  at first, the Nakanishi weight function, which then allows to 
restore the BS amplitude and the light front (LF) wave function. 

Other methods based on  the appropriate treatment of the singularities in the BS equation, provide the Minkowski 
solution directly, both for bound  and scattering states \cite{bs-long}. 

The straightforward numerical extrapolation of the solution from Euclidean to Minkowski space is  very unstable and has not achieved  any significant progress of practical interest. 
However, the above mentioned Nakanishi integral representation provides a more reliable method. 
It is valid for the Euclidean $\Phi_E$  as well as for the 
Minkowski $\Phi_M$ solutions and both solutions are expressed via one and the same Nakanishi weight function $g$. 
This integral representation, given in detail in the next section, can be symbolically written in the form: 
\begin{equation}\label{PhiNEg}
\Phi_{E}=K_E\,g  
\end{equation}
and 
\begin{equation}\label{PhiNMg}
\Phi_{M}=K_M\,g , 
\end{equation}
where $K_E$ and $K_M$ denote respectively the Nakanishi two-dimensional integral kernels in Euclidean and Minkowski  spaces  and g the weight function for the two-body bound state.

Solving the integral equation   (\ref{PhiNEg}) relative to $g$ and substituting the result into  relation 
(\ref{PhiNMg}),  one can in principle determine the Minkowski amplitude $\Phi_M$ starting with the Euclidean one  $\Phi_E$. 
This strategy seems  applicable since, in contrast to the direct  extrapolation  $\Phi_E\to\Phi_M$,  it uses the analytical properties 
of the BS amplitude which are implemented in the Nakanishi representation. 

A similar integral relation exist  expressing the LF wave function $\psi$ in terms of the Nakanishi weight function  $g$  \cite{bs1}: 
\begin{equation}\label{PsiLg}
Ã\psi_{LF}=L\,g  \end{equation}

One of the  most remarkable interests of the Nakanishi representation is that it constitutes a common root for the three different formal objects $(\Phi_E,\Phi_M,\psi_{LF})$
allowing to determine any of them once known any other, and thus providing a link
between two theoretical approaches -- Bethe-Salpeter and Light-Front Dynamics -- which are in principle and in practice quite different.

Since the LF wave function can be also found independently either by solving the corresponding equation \cite{cdkm} (see for instance \cite{MC_2000})
or by means of Quantum Field Theory (QFT) inspired approaches like the Discrete Light Cone Quantization  \cite{Brodsky98} or the Basis Light Front Quantization approach \cite{vary2016},  
one can pose the problem of finding the Nakanishi weight function from  equation (\ref{PsiLg}). 
This equation is only one-dimensional. Its solution is more simple, 
it requires matrices of  smaller dimension  and therefore it is more stable than 
the solution of the 2D  equation  (\ref{PhiNEg}).

It is  worth noticing that the same method  can be applied, far beyond 
the OBE dynamics, to the complete QFT dynamical content of the lattice calculations. Therein,  the  full  (not restricted to the OBE kernel)  Euclidean BS amplitude can be  
obtained and, via the Nakanishi integral,  the corresponding BS amplitude in Minkowski 
space can be calculated as well as the  observables. The Nakanishi representation was used in \cite{Craig} to calculate  the parton distribution amplitude for the  pion.

The first results of our research in this field were published in \cite{FCGK}. We have found that the solution of the 
integral equation (\ref{PhiNEg})  relative to $g$ was rather unstable. In order to find numerical solution, this integral 
equation was discretized using two quite different methods (Gauss quadratures and splines) and it resulted into a linear system. 
The instability of its the solution dramatically increases  with the matrix dimension.

It turned out that the equation (\ref{PhiNEg}) is  a Fredholm equation of the first kind, which mathematically
a classical example of an ill-posed problem. 
On the other hand, this equation has a unique solution. 
The challenge here is to use an appropriate method allowing  us to find this solution. 
These mathematical methods are developed and well known.  
Using them and knowing either $\Phi_E$ or $\psi_{LF}$  we aim to  extract $g$ and find from it  $\Phi_M$
and the corresponding observables.  
This will demonstrate that this procedure is feasible to get the solution. 

The appearance of mathematically well defined but numerically ill-posed problems  is not a rare exception.  
It is, for instance, manifested when using the Stieltjes and Lorentz integral transforms \cite{Efros_SJNP_1985,Efros:2007nq,Giussepina_EFB23_2016} 
to solve the scattering few-body problems with bound state boundary conditions
or when extracting the general parton distributions (GPD) from the experimental data \cite{Herve}.

In this work we will proceed according to the following  steps.  

In a first step, we will  solve numerically  equations (\ref{PhiNEg}) and   (\ref{PsiLg}) in a toy model where $\Phi_{E}$, $\psi_{LF}$ and $g$ are known analytically. 
The comparison of the numerical solutions for $g$ with the analytical one will tell us the reliability of our method in solving the one dimensional and two-dimensional Fredholm 
first kind equations. 

In a second step, we will go over a more realistic dynamical case. There,   $\Phi_E$ is obtained  by solving the Euclidean BS equation.
The LF wave function  is given by equation (\ref{PsiLg}), where $g$  is calculated  by  
using the methods introduced in \cite{bs1}, to solve the Minkowski BS  equation, and based on the LF projection and Nakanishi representation. 
In order to keep trace of its origin we will denote it  by $g_M$.

By solving  equations (\ref{PhiNEg})  and (\ref{PsiLg}), we find $g$  in two independent ways -- denoted respectively $g_E$ and $g_{LF}$ -- to be compared with each other as well as with $g_M$. 
Our methods should be consistent as far as all these weight functions are reasonably close to each other. 

This  wave function $\psi_{LF}$, found by projecting he BS amplitude in the light-front, and the $\psi_{LF}$ one  obtained from solving the LF equation, are practically indistinguishable from each other.
We find also the Minkowski BS amplitude $\Phi_M$ and using it, we calculate observables,  namely,
EM form factor and momentum distribution, represented by the LF wave function. We also calculate the 
form factor independently, expressing it via the LF wave function $\psi_{LF}$.  It turned out that the form factors 
calculated by these two ways are very close to each other and practically insensitive to instabilities of $g$'s remaining after their 
suppression by the method we use. That demonstrates 
that indeed, knowing the Euclidean BS amplitude and using the methods developed in the present paper, one can calculate electroweak observables.

In Sec. \ref{mod} we present the formulas for the Nakanishi  integral representation, both for the BS amplitude and the LF wave function. 
A change of variables (mapping) introduced  to simplify the integration domain is described in Sec. \ref{mapping}. 
The numerical method for solving the discretized equation is presented in Sec. \ref{solving}. 
The analytically solvable model, on which we will test the 
numerical solutions, comparing them with the analytical ones, is presented in Sec. \ref{analyt}.  
In Sec. \ref{stability} we study  the stability of these solutions. In Sec. \ref{num} we find numerically the Nakanishi weight function using as input the 
Euclidean BS amplitude and the LF wave function found from the OBE interaction kernel. These results are applied to calculate  EM form factors in Sec. \ref{ff}. 
Finally, Sec. \ref{concl} contains a genral discussion and the conclusions.

%%%%%%%%%%%%%%%%%%%%%%%%%%%%%%%
\section{Nakanishi representation}\label{mod}

The  BS amplitude  in Minkowski space $\Phi_M(k,p)$, for an $S$-wave two-body bound state with constituents masses $m$ ant total mass $M$, 
 depends  in the rest frame $p=(M,\vec{0})$ on two variables $k_v$ and $k_0$.
We represent the four-vector $k$ as  $k=(k_0,\vec{k})$ and denote $k_v=|\vec{k}|$. 

The Nakanishi representation for this amplitude reads \cite{nakanishi}:
%\begin{equation}\label{nak}
\begin{eqnarray}\label{nak}
&&\Phi_M(k_v,k_0) = 
\\
&&\int_{-1}^{1} dz' \int_{0}^{\infty} d\gamma'
\frac{g(\gamma', z')}{\left( \gamma' + \kappa^2 -k_0^2+k_v^2 -Mk_0z'-i\epsilon \right)^3}\, ,
\nonumber
\end{eqnarray}
%\end{equation}
where $\kappa^2 = m^2 - \frac{M^2}{4}$ and $g$ is the Nakanishi weight function. 
The power of the denominator is in fact an arbitrary integer  and we have chosen 3  for convenience.
This is just the equation which is symbolically written in  (\ref{PhiNMg}).

In the Euclidean space, after the replacement $k_0= ik_4$, this formula is rewritten as:
\begin{eqnarray}\label{nakE1}
&&\Phi_E(k_v,k_4) = 
\\
&& \int_{-1}^{1} dz' \int_{0}^{\infty} d\gamma'\frac{g(\gamma', z')}{\left( \gamma'  +k_4^2+k_v^2 + \kappa^2-iMk_4z' \right)^3},
\nonumber
\end{eqnarray}
which was represented symbolically in (\ref{PhiNEg}). 

One can also express through $g$ the LF wave function as \cite{bs1}:
\begin{small}
\begin{equation}
\label{lfwf1}
\psi_{LF}(\gamma,z)=\frac14\int_0^{\infty}\frac{(1-z^2)g(\gamma',z)d\gamma'}
{\Bigl[\gamma'+\gamma +z^2m^2+\kappa^2(1-z^2)\Bigr]^2}.
\end{equation}
\end{small}
This equation corresponds to (\ref{PsiLg}).
As  usually found in the literature, the LF wave function  $\psi_{LF}$ is considered as depending on  variables $k_{\perp}^2,x$. They are related to 
$\gamma,z$ by $\gamma=k_{\perp}^2,\; z=2x-1$ (see {\it e.g.} \cite{cdkm}).  

The Euclidean BS amplitude  $\Phi_E(k_v,k_4)$ can be easily found from the corresponding equation with a given kernel (OBE, for instance).  As mentioned,
it can be also found  in lattice calculations \cite{BS_Latt}, which are much  difficult  numerically but  include all  
the Quantum Field Theory dynamics.

We remind, that alternatively to eq. (\ref{PsiLg}) the LF wave function $\psi_{LF}$ can be found by solving a 2D equation in the LF dynamics \cite{cdkm}
or by using other QFT inspired methods like DLCQ \cite{Brodsky98} or BLFQ \cite{vary2016}. 

We will  therefore assume that  one of the functions $\Phi_E$  or $\psi_{LF}$ is known and can be used as input for solving  equations (\ref{nakE1}) or  (\ref{lfwf1})    
relative to Nakanishi weight function  $g$.  
Once obtained in this way, $g$ can be used to calculate any of the remaining quantities  in the triplet $(\psi,\Phi_E,\Phi_M$) and  corresponding observables. 

%%%%%%%%%%%%%%%%%%%%%%%%%%%%%%%%%%%%%%
\section{Mapping}\label{mapping}

Though,  the time-like momentum variable varies in the range $-\infty < k_4 < \infty$, we can reduce the problem to the half-interval $0 < k_4 < \infty$ and real arithmetic, 
by assuming $\Phi_E(k_v,-k_4) = \Phi_E(k_v,k_4)$.
Furthermore, we will make in Eq. (\ref{nakE1}) the following mapping   
$$
0< \gamma',k_v,k_4 < \infty\quad\to\quad 0 < x' ,x,z< 1
$$
by:
\begin{equation}\label{map}
\gamma'=\frac{x'}{1-x'},\quad k_v=\frac{x}{1-x}, \quad  k_4=\frac{z}{1-z}.
\end{equation}
Eq. (\ref{nakE1})  takes then the form:
\begin{small}
\begin{eqnarray}\label{nakE2}
&&\Phi_E(x,z) = 2 \int_{0}^{1} dx' \int_{0}^{1} dz' \;\frac{g(x', z')}{(1-x')^2}
\\
&\times&\mbox{Re}\left[\frac{x^2}{(1-x)^2}+\frac{z^2}{(1-z)^2}+\frac{x'}{1-x'}+ \kappa^2 -iM\frac{z z'}{1-z} \right]^{-3}.
\nonumber
\end{eqnarray}
\end{small}
to be solved in the compact domain $[0,1]\times[0,1]$.
The factor $\frac{1}{(1-x')^2}$ is the Jacobian.  We rewrite this equation as:
\begin{equation}\label{nakE2b}
\Phi_E(x,z) =  \int_{0}^{1} dx' \int_{0}^{1} dz'\, K(x,z;x',z')g(x',z')
\end{equation}
where 
\begin{small}
\begin{eqnarray}\label{K}
&&K(x,z;x',z')=\frac{2}{(1-x')^2}
\\
&\times&\mbox{Re}\left[\frac{x^2}{(1-x)^2}+\frac{z^2}{(1-z)^2}
+\frac{x'}{1-x'}+ \kappa^2 -iM\frac{z z'}{1-z} \right]^{-3}.
\nonumber
\end{eqnarray}
\end{small}
For the normal solutions $g(x',-z')=g(x',z')$ that comes from the symmetry 
of $\Phi_E$ with respect to $k_4$, which separates out the abnormal solutions 
for the two identical boson case.

After introducing in  Eq. (\ref{lfwf1}) for the variables $\gamma,\gamma'$ the mapping defined in (\ref{map}), we get that: 
\begin{equation}
\label{eq1}
\psi_{LF}(x,z)=\int_0^{1} dx'\, L(x,x';z)g(x',z)
\end{equation}
where 
\begin{eqnarray}\label{L}
L(x,x';z)&=&\frac{(1-z^2)}{4(1-x')^2}
\\
&\times&{\left(\frac{x}{1-x}+\frac{x'}{1-x'}+z^2m^2+\kappa^2(1-z^2)\right)^{-2}} .
\nonumber
\end{eqnarray}
Here 
$z$ plays the role of a parameter.

%%%%%%%%%%%%%%%%%%%%%%%%%%%%
\section{Solving equations (\ref{nakE2b}) and (\ref{eq1})}\label{solving}

We will look for the solution of Eq. (\ref{nakE2b})  by expanding it on  Gegenbauer polynomials in both variables:
\begin{equation}\label{g2}
g(x,z)=\sum_{i,j=1}^{N_x,N_z} c_{ij}G_{i-1}(x)G_{j-1}(z)
\end{equation}
where
\begin{equation}\label{Geg}
G_n(x)=\sqrt{2n+1}\, C^{\left(\frac{1}{2}\right)}_n(2x-1),
\end{equation}
$ C^{\left(\frac{1}{2}\right)}_n(2x-1)$ is a standard (non-normalized) Gegenbauer polynomial.
Whereas, the polynomial $G_n(x)$ are orthonormalized:
$$
\int_0^1 dx\, G_n(x) G_{n'}(x)=\delta_{nn'}
$$
We substitute $g(z,x)$ from (\ref{g2}) in Eq. (\ref{nakE2b}), calculate the integrals numerically and
validate  the equation in $N=N_x\times N_z$ discrete points $(x_i,z_j)$, with \mbox{$i=1,\ldots, N_x$} and \mbox{$j=1,\ldots, N_z$}. We chose as validation points $x_i$ ($z_j$) 
the $N_x$ ($N_z$) Gauss points  in the interval $0<x<1$.
Eq. (\ref{nakE2b}) transforms into the following linear system:
\begin{equation}
\label{eq1b}
\Phi_{ij}=\sum_{i'j'}^{N_z, N_x} K_{ij}^{i'j'}c_{i'j'}
\end{equation}
where
\begin{eqnarray}
\label{Kij}
\Phi_{ij}&=&\Phi_E(x_i,z_j),
\\
K^{i'j'}_{ij} &=&\int_0^1 dz'\int_0^1 dx K(x_i,z_j;z',x')G_{i'-1}(x')G_{j'-1}(z')
\nonumber
\end{eqnarray}

Eq. (\ref{eq1b}) is a $N\times N$  system of linear equations of the type $B=A\,C$ with the inhomogeneous
term given by the euclidean BS amplitude $B\equiv \Phi_{ij} $ and as unknowns the array $C\equiv c_{ij}$ of coefficients of the expansion  (\ref{g2}). 
The solution of this system  $C=A^{-1}\,B$, will provide the coefficients $c_{ij}$  and by this the Nakanishi weight function $g(x,z)$ at any point.

The solution of Eq. (\ref{eq1}) is found similarly. However, since $z$ is a parameter,  the solution is found for a fixed $z$ and its decomposition is one-dimensional:
\begin{equation}\label{g1}
g(x,z)=\sum_{i=1}^N c_iG_{i-1}(x)
\end{equation}
We substitute  $g$ in Eq. (\ref{eq1}), calculate the integral numerically and
validate  the  equation in the $N$  Gaussian quadrature points $\{x_j\}$ in the interval $0<x<1$. 
Equation (\ref{eq1}) transforms into the inhomogeneous  linear system (the parameter $z$ is omitted):
\begin{equation}
\label{eq1c}
\psi_{LF}(x_j)=\sum_{i=1}^N L_{ji}\,c_i,
\end{equation}
where
\begin{equation}
\label{Lij}
L_{ji} =\int_0^1 dx'\, L(x_j,x';z)\,G_{i-1}(x'),
\end{equation} 
with $L(x,x';z)$ defined in (\ref{L}).
For a given $z$, the system (\ref{eq1c})  of $N$ linear equations is solved, and once determined the coefficients $c_i$, Eq. (\ref{g1}) provides the solution  $g(x,z)$ everywhere.

Concerning the number of points  $N$ used in the discretization of the integral equations
there exists a  "plateau of stability", corresponding to an optimal value of $N$. 
For  a small  values of $N$, the accuracy is not enough but, as we will see below, by increasing $N$ the solution  becomes oscillatory and  unstable.
This is just a manifestation of above mentioned 
fact that the equations (\ref{nakE2b}) and (\ref{eq1}) represent both  a Fredholm integral equation of the first kind which is 
a classical example of an ill-posed problem. Their kernels are quadratically integrable what ensures the existence and uniqueness of the solution. 
However, direct numerical methods do not allow to find the solution in practice, since they  lead to unstable results. To avoid instability, 
one can, of course, keep $N$ small enough. However, for small $N$ the expansions (\ref{g2}) and (\ref{g1}) give a very crude reproduction of the unknown $g$. 
Therefore, to find solution, we will use a special mathematical method -- the Tikhonov regularization  (TRM) \protect{\cite{Tikhonov}}.

Note that another method (Maximum Entropy Method)  was  recently proposed \cite{MEM}   to solve Eq. (\ref{nakE1}) and successfully applied, at least in the case
of monotonic $g$'s.

We will follow here the standard and straightforward way explained above: by  discretization of
the integral equation we turn it into  a matrix equation 
and solve it by inverting the matrix, however, regularizing the inversion problem. 
The TRM \cite{Tikhonov}  allows to find a stable solution for sufficiently large $N$. 
Namely, following to \cite{Tikhonov}, we will  solve an approximate minimization problem, i.e., we will find $C$ providing the minimum of:
$$
\parallel\, A\,C\,-\,B\,\parallel\, .
$$
In the normal form,  that corresponds to the    
replacement of the equation  $A\,C\,=\,B$ by the regularized one  $A^{\dag}AC+\epsilon C=A^{\dag}B$ with $\epsilon\ll 1$. The solution of this equation reads: 
\begin{equation}
\label{reg}
C_{\epsilon}=(A^{\dag}A+\epsilon I)^{-1}A^{\dag}B,
\end{equation}
where $I$ is the identity matrix. For small  $\epsilon$ (but not for infinitesimal)  the solution $C_{\epsilon}$ of this equation is very close to the solution of  the original equation $AC=B$. However, the solution  of Eq. (\ref{reg}) is much more stable than the solution  of $AC=B$.

 In our previous work \cite{FCGK} we also used a regularization procedure, but in a more naive form. The equation $AC=B$ was replaced not by 
 $A^{\dag}AC+\epsilon C=A^{\dag}B$, 
but by $(A+\epsilon I) C=B$. Then, instead of the solution (\ref{reg}) we got
\begin{equation}\label{reg1}
C_{\epsilon}=(A+\epsilon I)^{-1}B.
\end{equation}
This increases the stability but not so strongly like (\ref{reg}), as we are going to illustrate in what follows.

%%%%%%%%%%%%%%%%%%%%%%%%%
 \section{Analytically solvable model}\label{analyt}

As the first step, we will check the methods and study the  appearance of numerical instabilities by solving the simplest one-dimensional equation (\ref{lfwf1}) in a model where the functions $\psi_{LF}$ and $g$ are known analytically. We will compare the numerical solution with the analytical one. In  equation (\ref{lfwf1}) the values $z,m$ and $\kappa$ are the parameters which can be chosen arbitrary (keeping the kernel non-singular). We can also put an arbitrary factor at the front of the integral in r.h.-side, which only changes the normalization of $\psi_{LF}$. 
Using this freedom, we choose $z^2m^2+\kappa^2(1-z^2)=1$ and replace the factor $\frac{1}{4}(1-z^2)$ also by 1.  Then the equation (\ref{lfwf1}) obtains the form:
\begin{equation}
\label{eq2}
\psi_{LF}(\gamma)=\int_0^{\infty}\frac{g(\gamma')d\gamma'}
{(\gamma+\gamma' +1)^2}.
\end{equation}
As a solution we take:
\begin{equation}
\label{eq3}
g(\gamma)=\frac{1}{(1+\gamma)^2}
\end{equation}
Its  substitution it into (\ref{eq2}) provides  the l.h.-side:
\begin{equation}
\label{eq4}
\psi_{LF}(\gamma)=\frac{1}{\gamma^3}\left[\frac{\gamma(2+\gamma)}{(1+\gamma)}-2\log(1+\gamma)\right]
\end{equation}

In the mapping variables the equation (\ref{eq2}) is rewritten as:
\begin{equation}
\label{eq1a}
\psi_{LF}(x)=\int_0^1 L_1(x,x')g(x')dx',
\end{equation}
where the kernel $L_1$, the inhomogeneous term $\psi_{LF}$ and the solution $g$ read:
\begin{eqnarray}
\psi_{LF}(x)&=&\Bigl[(2-x)x+2(1-x)\log(1-x)\Bigr]
\nonumber\\
&\times&\frac{(1-x)^2}{x^3},
\label{psix}\\
L_1(x,x')&=&\frac{1}{\left[\frac{x}{1-x}+\frac{x'}{1-x'}+1\right]^2}\frac{1}{(1-x')^2}
\nonumber\\
&=&\frac{(1-x)^2}{(1-xx')^2},
\label{L1x}
\\
g(x)&=&(1-x)^2.
\label{gx}
\end{eqnarray}
We will consider $\psi_{LF}$, given by Eq. (\ref{psix}) as input, solve numerically, Eq. (\ref{eq1a})  in the form of the decomposition (\ref{g1})
and compare the  solution with the analytical $g$ given in (\ref{gx}). 

\begin{figure}[thb!]
\begin{center}
\includegraphics[width=0.5\textwidth]{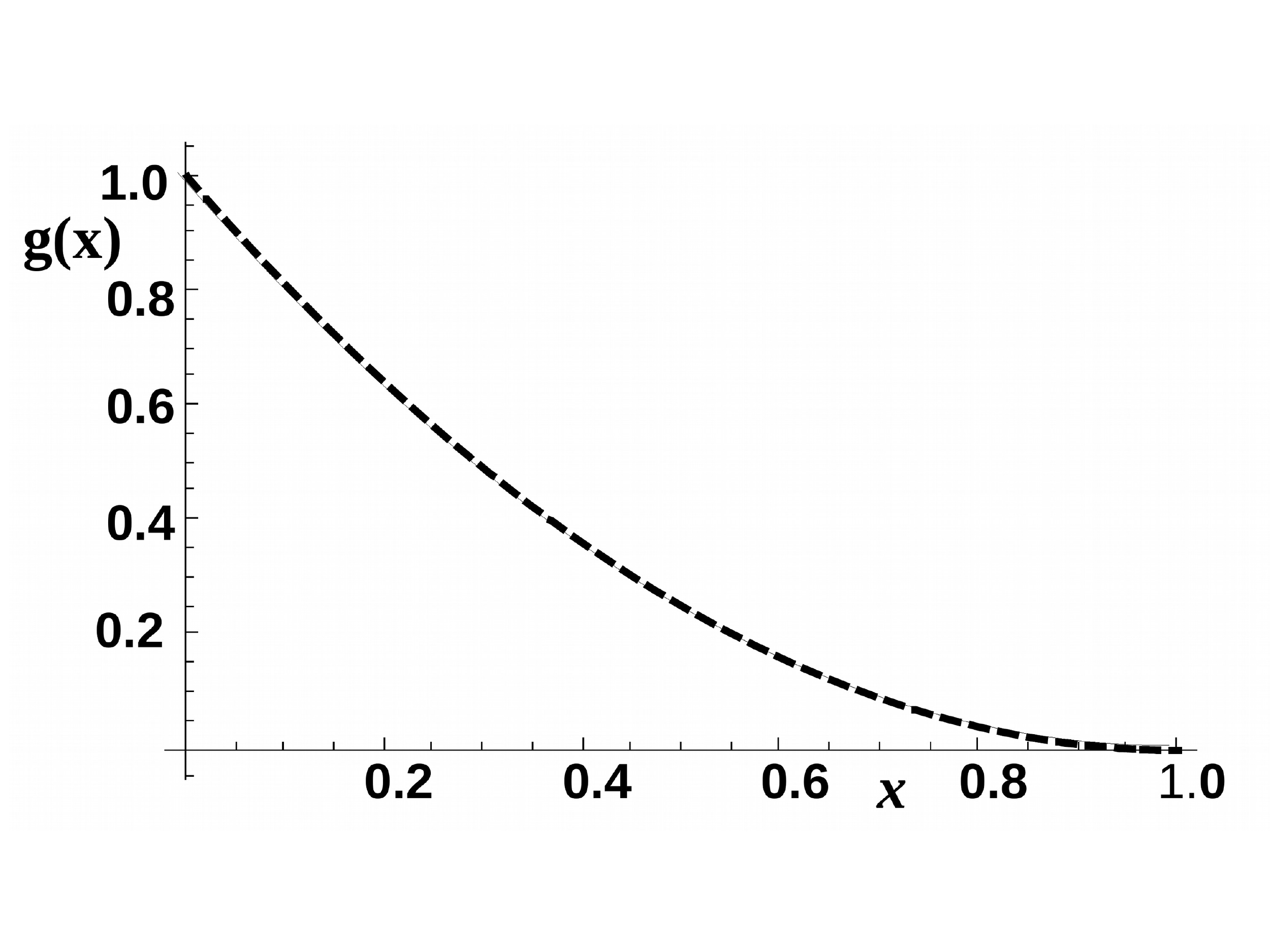}
\end{center}
\vspace{-1.cm}
\caption{The numerical solution of Eq. (\ref{eq1a}) for $g(x)$ (dashed  curve), found in the form of Eq. (\ref{g1}), for the discretization rank $N=11$ with  $\epsilon=0$ in (\ref{reg1}) in comparison to the exact solution $g(x)=(1-x)^2$ (not-distinguishable from the dashed curve).}\label{gN11eps0}
\end{figure}

%%%%%%%%%%%%%%%%%%%%%%%%%%%%%%%%%%%%%%%%%%%
\section{Studying stability}\label{stability}

In our precedent paper \cite{FCGK} we have found that the inversion of the Nakanishi integral is rather unstable 
relative to the increase of the number of points $N$, using the regularization method given by Eq. (\ref{reg1}).
Therefore, solving Eq. (\ref{eq1c}), we will first study  the onset of instability  
and its suppression by the Tikhonov regularization (\ref{reg}). 

\begin{figure}[ht!]
\begin{center}
\includegraphics[width=0.5\textwidth]{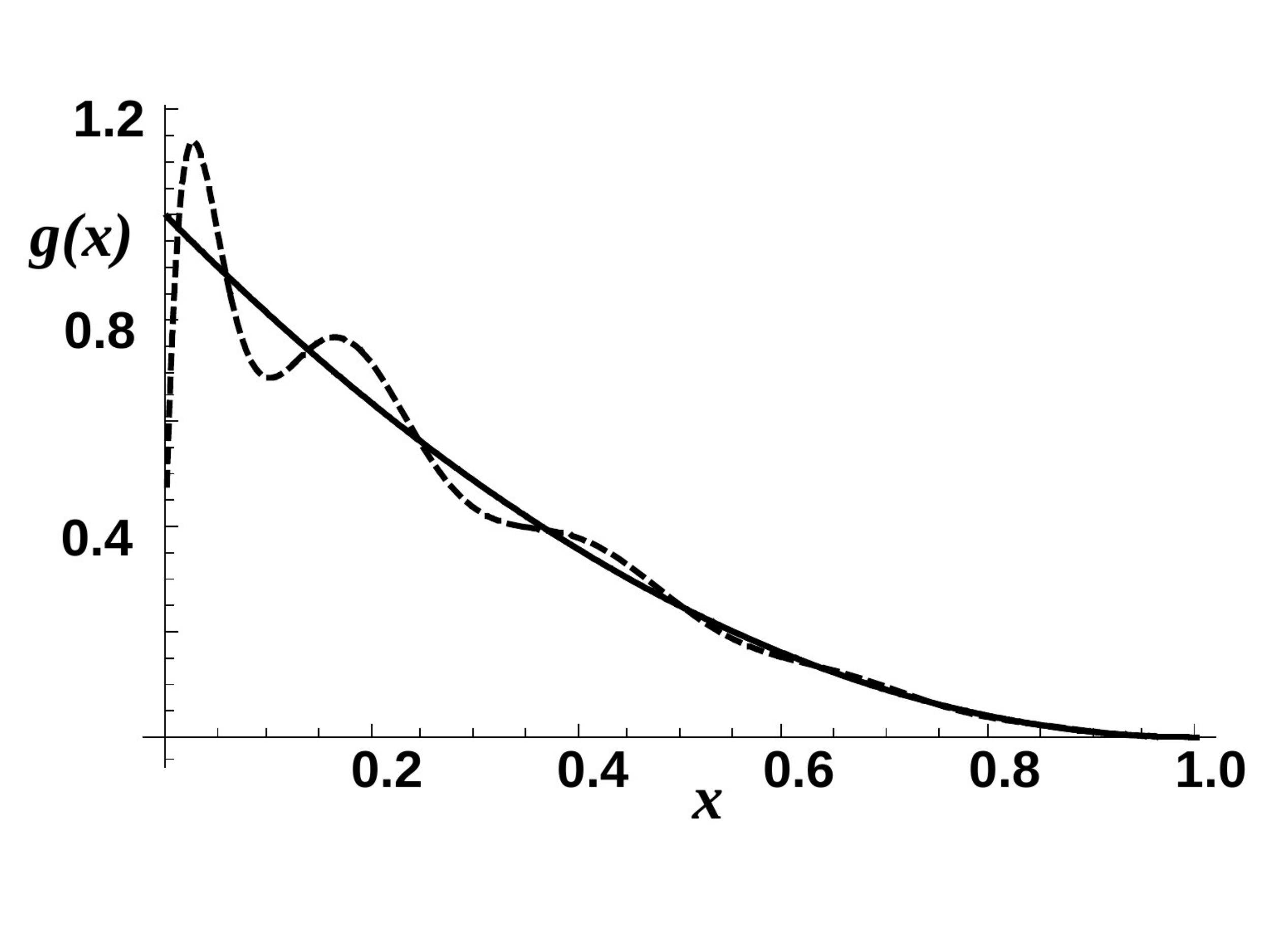}
\end{center}
\vspace{-1.cm}
\caption{The same as in Fig. \ref{gN11eps0} but for $N=14$, $\epsilon=0$.}\label{gN14eps0}
\end{figure}

The solution of the non-regularized equation (\ref{eq1c}), by using Eq. (\ref{reg1}) with $\epsilon=0$ and $N=11$,  
is shown in  Fig.~\ref{gN11eps0}.
It coincides, within the thickness of the lines, with the exact one, given by Eq. (\ref{gx}).  
However, when increasing $N$ up to $N=14$, 
we can observe  (Fig.~\ref{gN14eps0}) the onset of instability: the numerical solution $g(x)$  becomes oscillating around the exact solution. 
Note also that the determinant of $A$ is very small and it quickly decreases when $N$ increases. 
For $N=11$, det$(A)\sim 10^{-35}$ while for $N=14$  one has det$(A)\sim 10^{-59}$.

\vspace{-0cm}
\begin{figure}[ht!]
\centering
\includegraphics[width=0.5\textwidth]{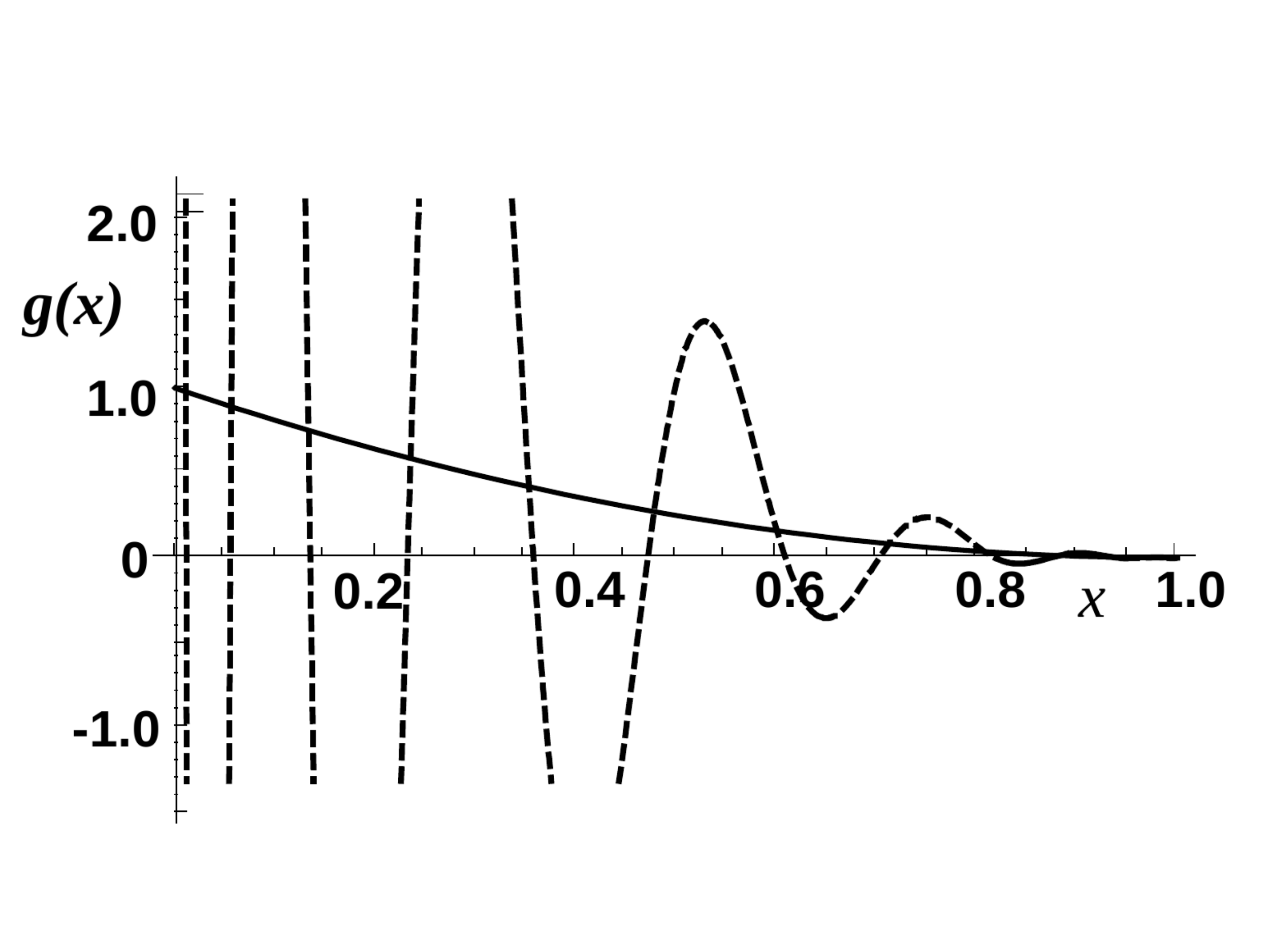}
\vspace{-1cm}
\caption{The same as in Fig. \ref{gN11eps0} but for $N=16$, $\epsilon=0$.}\label{gN16eps0}
\end{figure}
\vspace{-0cm}

Further increase of $N$ results in extremely strong oscillations. 
The solution for $N=16$ and $\epsilon=0$,  strongly oscillates and  dramatically differs 
from the exact one, as it can be seen in  Fig.~\ref{gN16eps0} . 
For $N=32$ and $\epsilon=0$, the solution  oscillates even more strongly and is wrong by orders of magnitude;
we find for instance $g(0)\approx -10^3$ instead of $g(0)=1$ according to the analytic result.

\begin{figure}[ht!]
\centering
\includegraphics[width=0.5\textwidth]{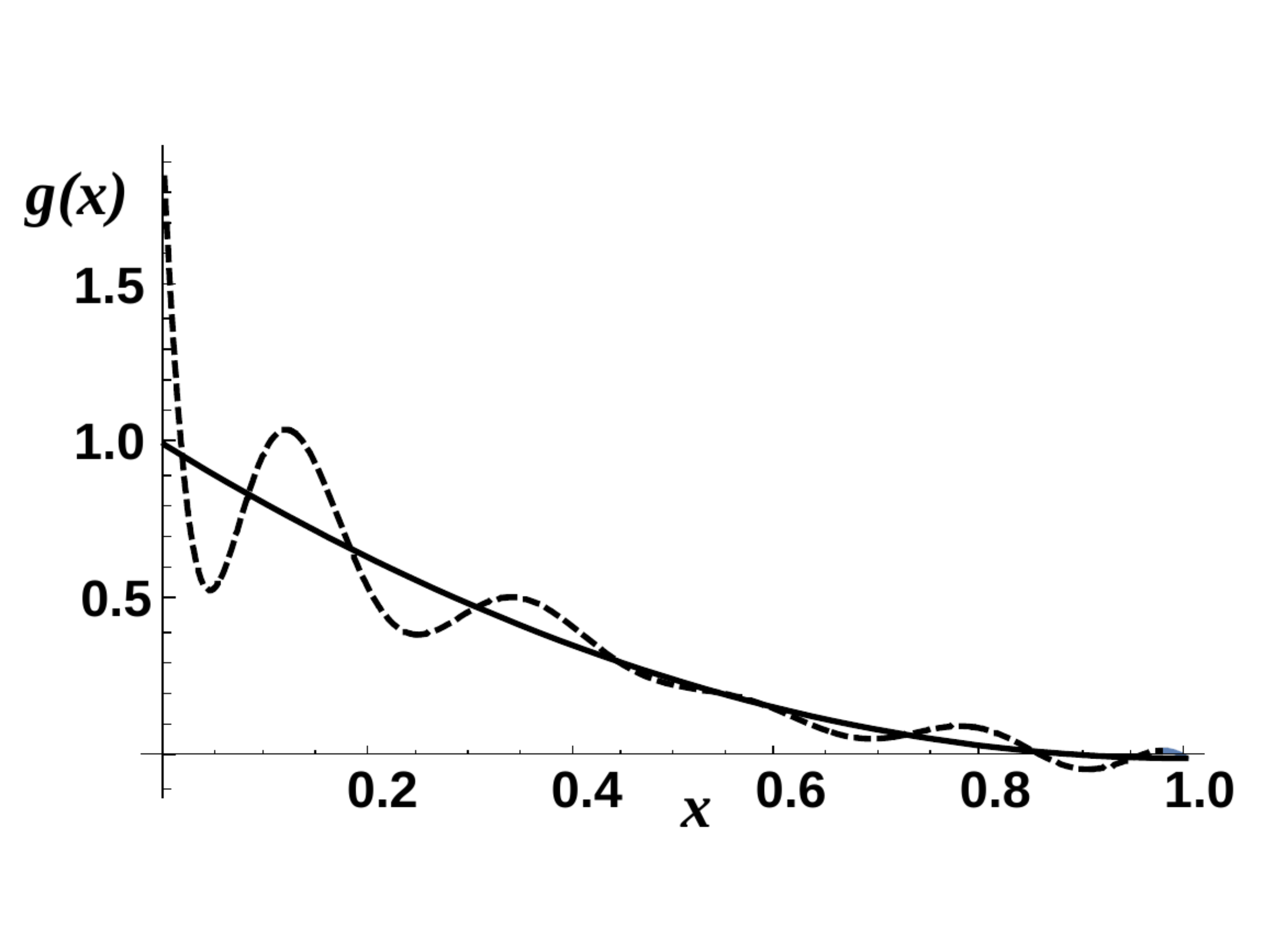}
\vspace{-1cm}
\caption{The same as in Figs. \ref{gN11eps0}, \ref{gN14eps0}, \ref{gN16eps0},  but with Tikhonov regularization, for $N=16$ and $\epsilon=10^{-18}$.}
\label{gN16eps-18}
\end{figure}

\begin{figure}[ht!]
\centering
\includegraphics[width=0.5\textwidth]{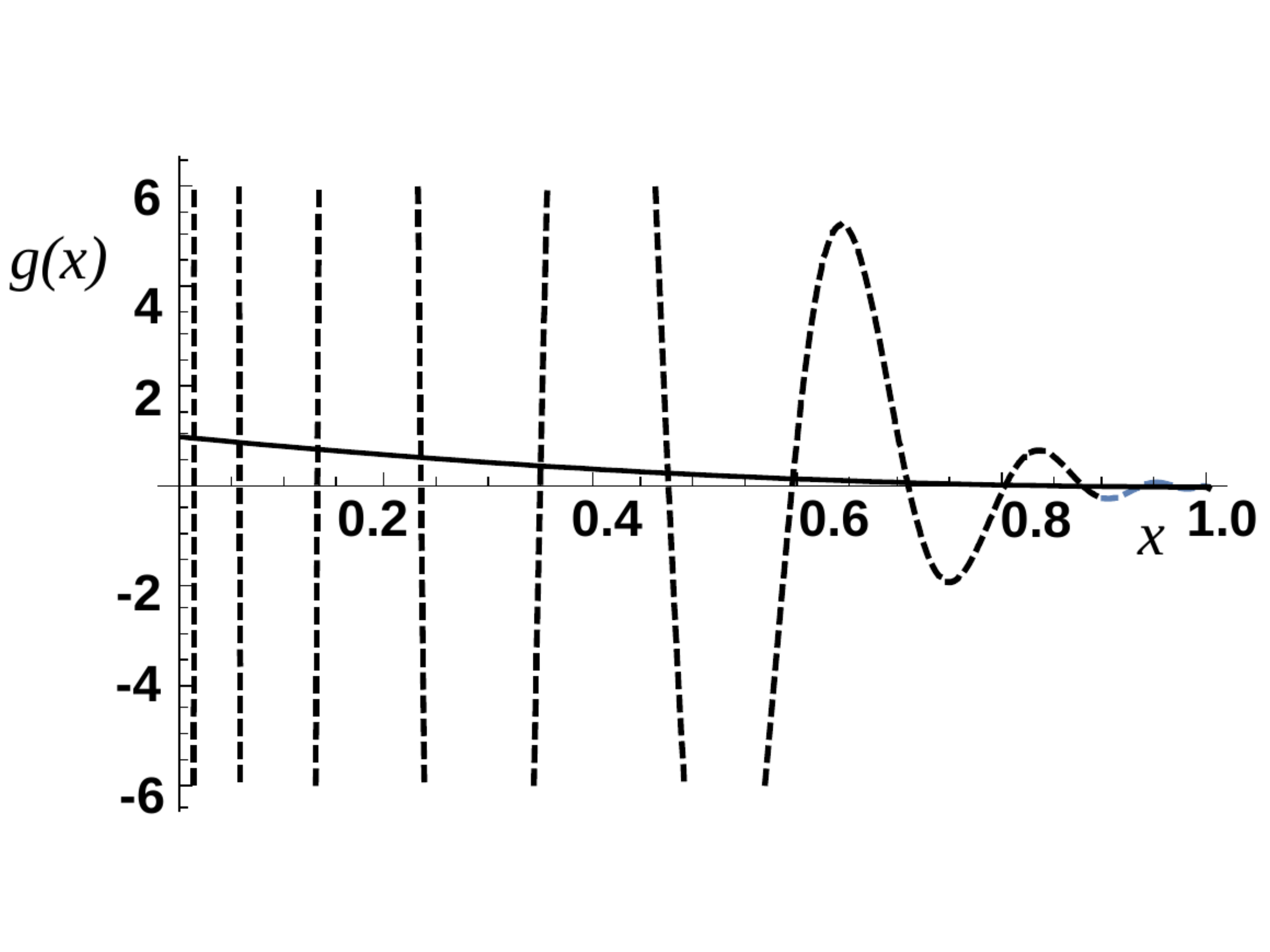}
\vspace{-1cm}
\caption{The same as in Fig. \ref{gN16eps-18} but with the regularization by Eq. (\ref{reg1}), $N=16$ and  $\epsilon=10^{-10}$.}\label{gN16eps-10}
\end{figure}

Now let us solve the corresponding equation by using TRM,  Eq. (\ref{reg}). For $N=16$ and $\epsilon=10^{-10}$, we found that 
the oscillations completely disappear. The numerical solution coincides with the exact one 
within precision better than 1\%, similarly to what we observe in the  figure~\ref{gN11eps0}. 
To avoid repetition, we do not show the corresponding figure.
For much smaller $\epsilon=10^{-18}$ with $N=16$, the solution found by TRM, Eq. (\ref{reg}), is
shown in Fig.~\ref{gN16eps-18}. It has some oscillations, which are strongly enhanced in the
case with regularization (\ref{reg1}) and  $\epsilon=10^{-10}$.  The same happens with 
the regularized  solution for $N=14$, $\epsilon=10^{-10}$ (not shown), when we replace the Tikhonov regularization (\ref{reg})  by (\ref{reg1}).

Our  study  shows that using the TRM method  there is  almost  no $\epsilon$ dependence of the results   in a rather wide limits. 
However, when $\epsilon$ is taken relatively large (e.g. $N=16$, $\epsilon=10^{-4}$), 
the numerical solution can sensibly differs from the exact one. 
On the other hand, as it was mentioned,  when using too small   
$\epsilon$  one recovers the problem of oscillations, as it is seen in  Fig.~\ref{gN16eps-18}  ($N=16$ and $\epsilon=10^{-18}$).
For  $N=16,\epsilon=10^{-17}$  one can observe the first, yet weak signs of oscillations.  Hence,
for $N=16$, the stability (absence of oscillations) and insensitivity to the value of $\epsilon\neq 0$ are valid in rather large interval 
$\epsilon=10^{-4}\div 10^{-17}$.

 The regularized solution for $N=32,\epsilon=10^{-10}$, found by Eq. (\ref{reg}) and TRM, again coincides with the exact one within thickness of lines.
We do not show it since the curves are the same as in Fig.~\ref{gN11eps0}.
For $N=32$,  the oscillations appear at  $\epsilon\leq 10^{-16}$.

For higher $N$ the situation is similar. Like in the case $N=32$, the solution for $N=64,\epsilon=0$, found by $C=A^{-1}B$ strongly oscillates and it strongly differs from the exact one, even stronger than in Fig.~\ref{gN16eps0}. The calculated value $g(0)\approx -10^7$ instead of $g(0)=1$ is again completely wrong.
The regularized solution for $N=64,\epsilon=10^{-10}$, found by Eq. (\ref{reg})   also coincides with the exact one within thickness of lines,
like it is in  Fig.~\ref{gN11eps0}.
However, the  strong oscillations appear earlier, at  $\epsilon\leq 10^{-13}$ (in contrast to   $\epsilon\leq 10^{-16}$ for $N=32$).

 We compare  now the two ways of regularization given by Eqs. (\ref{reg1}) 
and (\ref{reg}). In Fig.~\ref{gN16eps-10} the solution given by Eq. (\ref{reg1}) for $N=16$, 
$\epsilon=10^{-10}$ is shown. It oscillates, while, the solution found
 by using TRM, Eq.(\ref{reg}), with same N and $\epsilon$ (not shown) coincides with the analytical one. 
 As mentioned above, the solution for $N=16$ and $\epsilon=10^{-18}$ (Fig.~\ref{gN16eps-18}) reveals moderate oscillations, which disappear at $\epsilon>10^{-17}$.

The reason which makes the determinant very small and turns the solution of the linear system in an ill-conditioned problem is the presence of very small eigenvalues of the kernel of the Fredholm integral equation (\ref{eq1a}).
As the dimension of the matrix increases, more small 
eigenvalues are present, and the eigenstates are oscillating functions, that mixes with the solution obtained within 
a numerical accuracy. These contributions to the solution are oscillatory, building the pattern seen in Figs.~\ref{gN16eps0} and \ref{gN16eps-10}, 
by increasing dimension of the matrix equation. The amplitude of the eigenvector contribution to the solution increases
 as the eigenvalue decreases, making the oscillations of the solution  divergent. The regularization
by $\epsilon$ cuts the contribution of the small eigenvalues, and to some extend  the numerical stability 
can be found, at the expense of a finite $\epsilon$. Both regularization  methods from Eq. (\ref{reg}) (Tikhonov method) and (\ref{reg1}) 
improve the numerical solution. The Tikhonov regularization with the reduction to the normal form provides larger eigenvalues as compared to
(\ref{reg1}), as it is evident from the stability analysis that in practice allows much smaller $\epsilon$'s before considerable 
oscillations of the solution appear. See for example Fig.~\ref{gN16eps-10}, where the method (\ref{reg1}) provides huge oscillations.
Whereas the TRM allows $\epsilon$ small as $10^{-18}$ compared with the solution with  $\epsilon=10^{-10}$.

The calculations presented in this section demonstrate that the Nakanishi representation, at least,  for the LF wave function, Eq.   (\ref{lfwf1}), can be 
indeed inverted numerically, that is, be solved relative to the Nakanishi weight function $g$. 
Though this representation, considered as an  equation for $g$, is an ill-posed problem,  special methods, in particular those based on the Tikhonov regularization (\ref{reg}),  allows to solve it. 

In the next section we will show that the Nakanishi representation  (\ref{nakE1}) for the Euclidean BS amplitude can be also inverted. 
This will be shown not in a toy model, but for the BS solution with a OBE kernel.

%%%%%%%%%%%%%%%%%%%
\section{OBE interaction}\label{num}

Let us now  consider the  dynamical case of  two spinless particle of unit mass ($m=1$), interacting via OBE kernel with exchanged boson mass $\mu=0.5$, and forming a  bound state of total mass $M=1.0$.

All the necessary solutions in this model  -- LF wave function $\psi_{LF}$,  Euclidean BS amplitude $\Phi_E$  
and Nakanishi weight function  $g$ --    have been computed by solving the corresponding equations. 
The weight function $g$ was found  independently from the LF wave function and BS solutions  
by solving the equation  derived in \cite{bs1,fsv_2014} with the same OBE kernel.  We will use 
for  $\psi_{LF}$ and $g$  the results \cite{fsv_2014,Gianni} and for $\Phi_E$ --  our own calculations.  % the results          

We will take profit  from  the simplicity of the Euclidean solution and  extract the Nakanishi weight function 
$g_E$ from  $\Phi_E$ by inverting  the Nakanishi representation in Euclidean space (\ref{nakE1})
via the Tikhonov regularization method and compare it with $g$  found from an equation derived from the 
Minkowski space BS equation \cite{fsv_2014,Gianni}. 
The quality of the solution $g_E$ will be checked by computing the  LF wave function 
$\psi_{LF}$ and comparing it with  $\psi_{LF}$ found via $g$ provided by \cite{Gianni}. Schematically the above extraction procedure is represented as $\Phi_E \to g_E\to \psi_{LF}$. 
Another possibility to extract $g$ is $\psi_{LF}\to g_{LF}\to \Phi_E$. Namely, starting with the LF wave function,  get $ g_{LF}$  and calculate with it 
the Euclidean $\Phi_E$ and  compare  this result with the initial $\Phi_E$. We  will  also compare  the EM form factors 
calculated initially via $\psi_{LF}$ and, independently, via $\Phi_M$ and finally expressed via $g$.
We will see that the observables are almost insensitive to the residual uncertainties 
   in either $g_E$ or $g_{LF}$, which survive after suppressing the instabilities by 
  the Tikhonov regularization (\ref{reg}). This ensures reliable results for the observables, when one uses 
  the Euclidean BS amplitude as input, calculates $g$ by inverting the Nakanishi integral and 
  then with the extracted $g$ goes to the observables.  

\begin{figure}[ht!]
\centering
\includegraphics[width=0.5\textwidth]{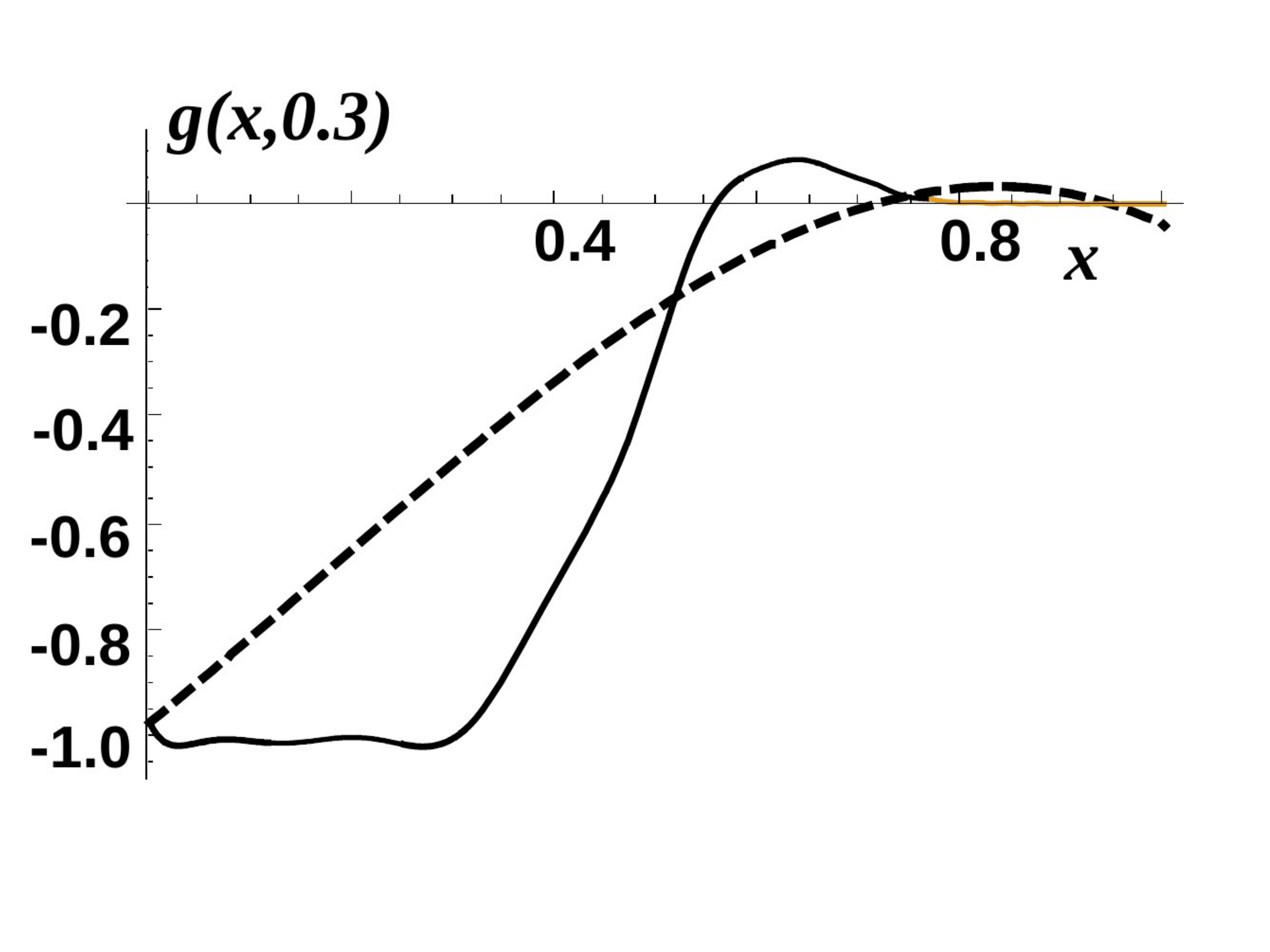}
\caption{The solution $g_E(x,\,z=0.3)$ (dashed) found solving Eq. (\ref{nakE2b}) with $N_z=2,N_x=4$ and $\epsilon=10^{-10}$, 
with $\Phi_E$, using the Tikhonov regularization method, compared to  the solution $g_{FSV}(x,\,z=0.3)$ 
\cite{fsv_2014,Gianni}  (solid).}
\label{fg42}
\end{figure}

The Nakanishi weight function $g_E$  obtained by inverting eq. (\ref{nakE2b}) with 
 $N_x=4,N_z=2$  and the TRM with $\epsilon=10^{-10}$,  is shown in   Fig.~\ref{fg42}  (dashed line). 
 Solid line represents the direct solution $g$    -- denoted below as $g_{FSV}$ -- 
computed in~\cite{fsv_2014,Gianni} by Frederico-Salm\`e-Viviani by solving a dynamical equation for $g$ and normalized in a different way than $\Phi_E$.  
To compare both solutions for $g$, we normalize $g_E$ so that the two solutions coincide (and equal to $-1$) at $x=0,\,z=0.3$. 

\begin{figure}[ht!]
\centering
\includegraphics[width=0.5\textwidth]{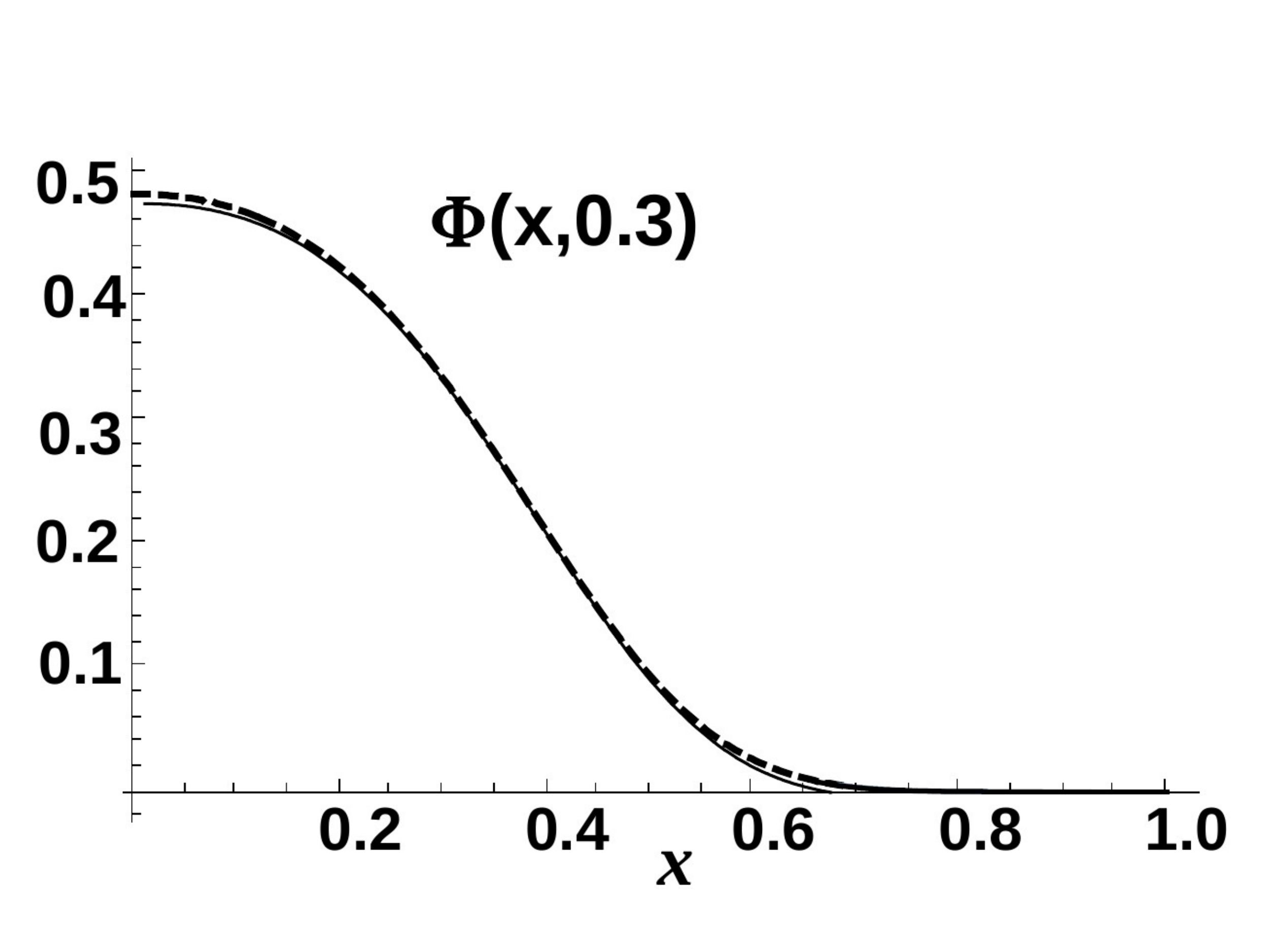}
\vspace{-.5cm}
\caption{The solid curve corresponds to the Euclidean BS amplitude 
$c_2\,\Phi_E(z=0.3,x)$ ($c_2=0.88$) calculated via the BS equation with the OBE kernel. 
The dashed curve corresponds to the BS amplitude  $\Phi_E(z=0.3,x)$
calculated via $g_E$ by Eq. (\ref{nakE2b}).}
\label{fphi42}
\end{figure}

To check the quality of $g_E$, solution of Eq. (\ref{nakE2b}), we substitute it in r.h.-side of Eq. (\ref{nakE2b}), 
calculate $\Phi_E$  and compare it with the input $\Phi_E$.  
The result is shown in  Fig.~\ref{fphi42}.  We display   
by the solid curve the input Euclidean BS amplitude $c_2 \, \Phi_E$ as a function of $x$ for a fixed value $z=0.3$.  The dashed curve corresponds to the BS amplitude  $\Phi_E(x,z=0.3)$ calculated via $g_E$ by Eq. (\ref{nakE2b}). Up to a factor  $c_2=0.88$ they are very close to each other.

\begin{figure}[h!]
\centering
\includegraphics[width=0.5\textwidth]{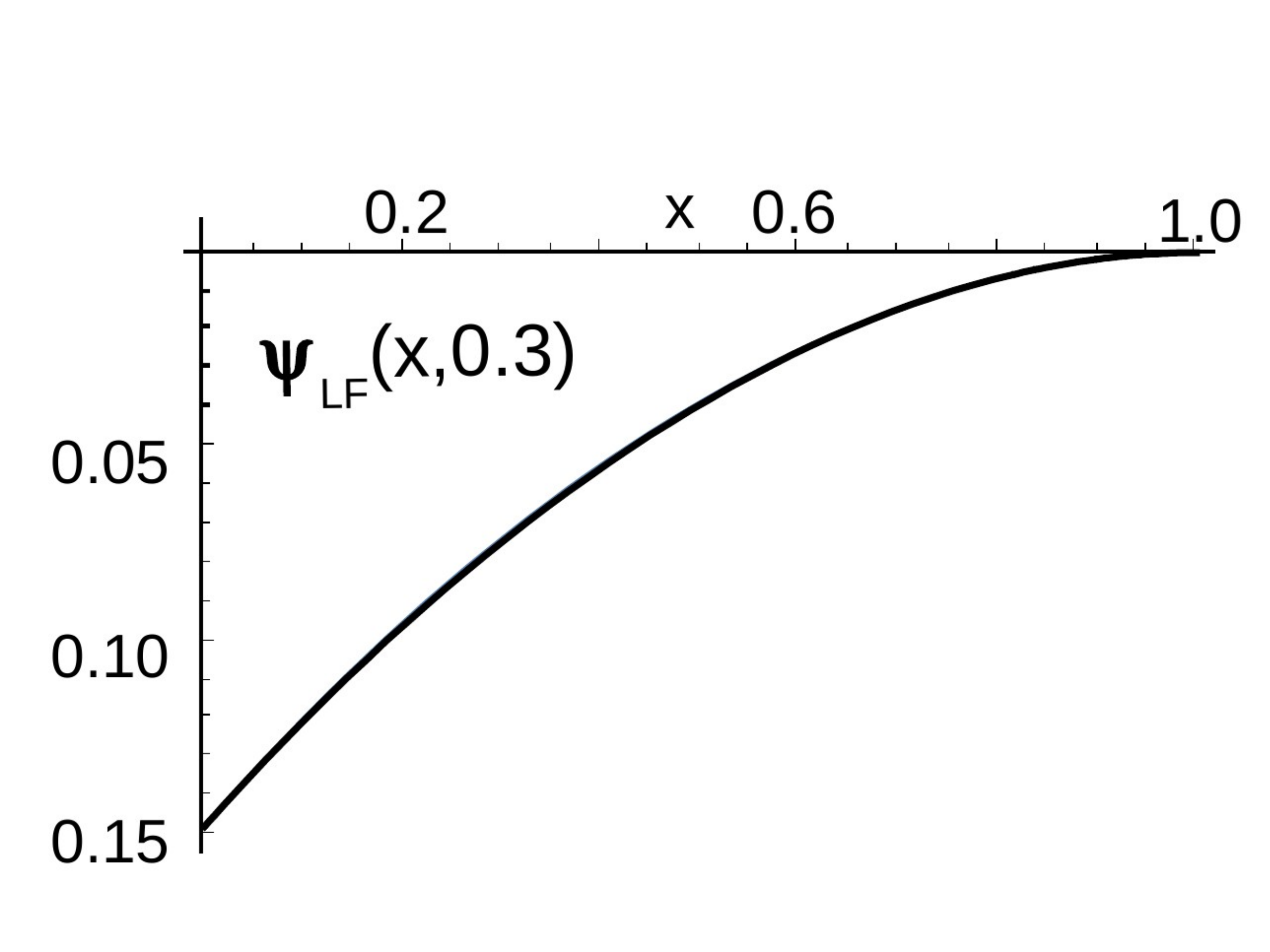}
\vspace{-0.5cm}
\caption{LF wave function calculated  by Eq. (\ref{eq1}) (multiplied by a normalization factor)
for both extracted $g_{LF}$ and $g_E$
compared to the actual results obtained with $g_{FSV}$ in \cite{fsv_2014,Gianni} from the Minkowski 
space solution of the BS equation. All the curves overlap with each other within the width of the lines.}
\label{fpsi42}
\end{figure}

The corresponding LF wave functions  calculated from $g_E$ and $g_{FSV}$ by using Eq. (\ref{eq1}) are shown in Fig.~\ref{fpsi42}. 
We multiply $\psi_{LF}$ calculated by Eq. (\ref{eq1}) 
 by a normalization factor.  Both $\psi_{LF}$'s well coincide with each other despite the difference between $g_E$ and $g_{FSV}$ seen in Fig.~\ref{fg42}.

\begin{figure}[h!]
\centering
\includegraphics[width=0.5\textwidth]{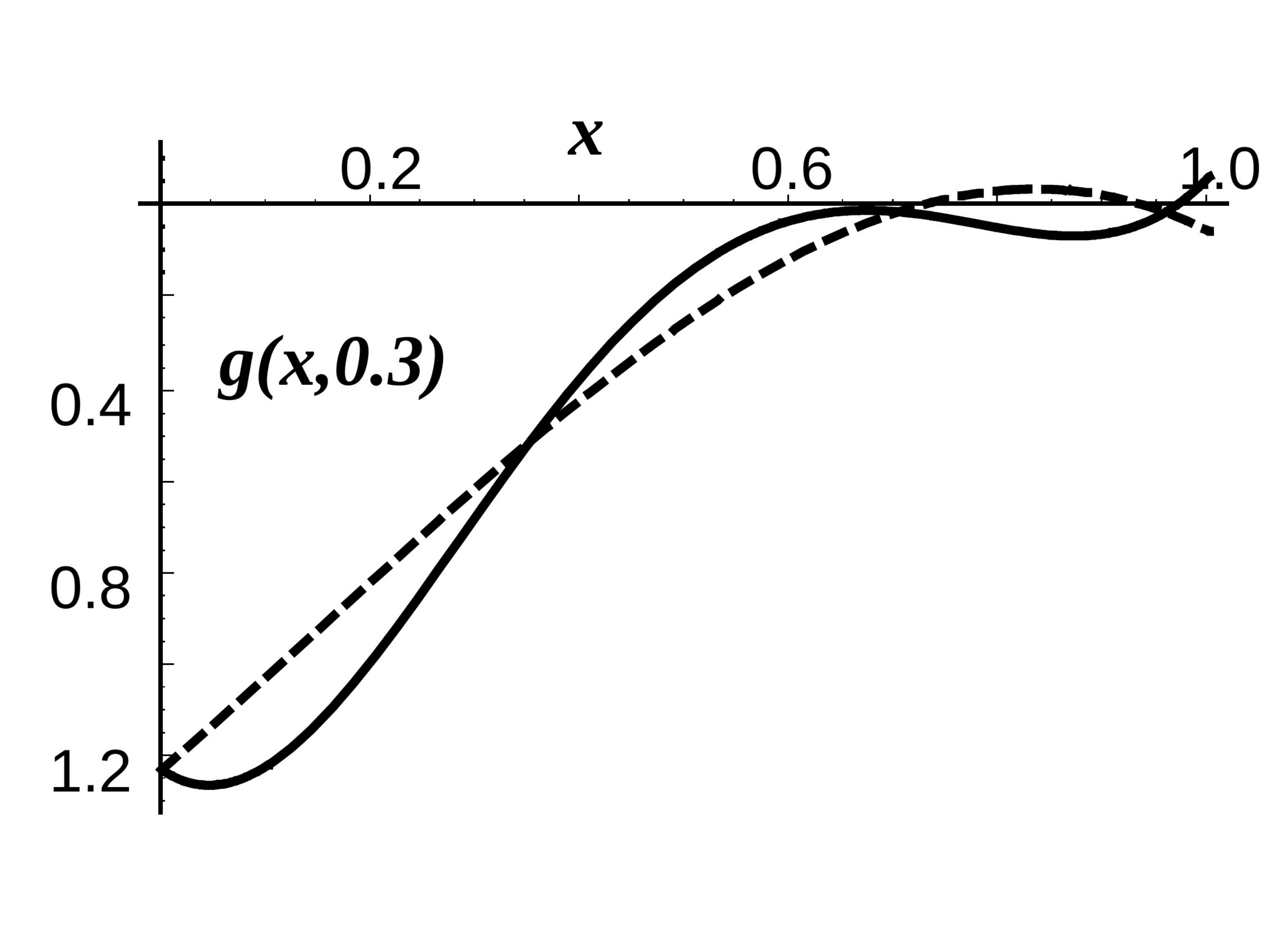}
\vspace{-1cm}
\caption{ The Nakanishi weight function $g_E(x,z=0.3)$ extracted from $\Phi_E$, with $N_z=2,\,N_x=4$ (dashed curve) and 
$N_x=5,\,N_z=3$ (solid curve) with $\epsilon=10^{-10}$ using TRM. }
\label{fg42a}
\end{figure}

The solution $g_E$ for $N_x=5$, $N_z=3$, $\epsilon=10^{-10}$ 
multiplied by a normalization factor,  is shown in Fig.~\ref{fg42a}.
For comparison, we add the solution for  $N_x=4$, $N_z=2$, $\epsilon=10^{-10}$, shown by dashed line  in Fig. \protect{\ref{fg42}}. In spite of the visible  difference between the dashed and solid curves, these two solutions give coinciding LF wave functions and the Euclidean BS amplitudes (after equivalent normalizations), similarly to ones  shown in Figs. \ref{fphi42} and \ref{fpsi42}.

 %%%%%%%%%%%%%%%%%%%%%%%%%%
\section{Calculating EM form factor}\label{ff}

Once we have the Nakanishi weight function $g$ by Eqs. (\ref{nak}) and (\ref{lfwf1}), we can find both
 the BS amplitude in Minkowski space and the LF wave function. The EM form factor can be expressed via both of them. In this way, we obtain two expressions
  in terms of  $g$. 
We will use both to calculate the form factor and we will compare the results.

The electromagnetic vertex is expressed in terms of the BS amplitude by:
\begin{eqnarray}\label{ffbs}
&&(p+p')^{\nu} F^{BSM}(Q^2) =  i\int \frac{d^4k}{(2\pi)^4}\,
(p+p'-2k)^{\nu}
\nonumber\\
&\times& (k^2-m^2)\Phi_M \left(\frac{1}{2}p -k;p\right)\Phi_M \left(\frac{1}{2}p'-k;p'\right) ,
\end{eqnarray}
where $Q^2=-q^2$, $q$ is the four-momentum transfer, and $p'=p+q$.
Substituting $\Phi_M$ in the form (\ref{nak}) and calculating the integral over $d^4k$, one gets
(see Eq. (13) from   \cite{ckm_2009}):
\begin{eqnarray}\label{ffM}
&&F^{BSM}(Q^2)=\frac{1}{2^7\pi^3 N_{BSM}}\int_0^\infty d\gamma \int_{-1}^1
dz\, g(\gamma,z)
\nonumber\\
&\times&  \int_0^\infty d\gamma' \int_{-1}^1 dz'\,
g(\gamma',z')\int_0^1 du\,u^2(1-u)^2 \frac{f_{num}}{f^4_{den}}
\end{eqnarray}
with
\begin{eqnarray*}
f_{num}&=&(6 \xi-5)m^2 +
           [\gamma' (1 - u) + \gamma u] (3 \xi-2)
\\
&+&2M^2 \xi(1-\xi) + \frac{1}{4} Q^2 (1 - u) u (1+z) (1+z')
\\
f_{den}&=& m^2 + \gamma' (1 - u) + \gamma u -
            M^2 (1 - \xi) \xi
           \\
&+& \frac{1}{4}Q^2 (1 - u) u (1+z) (1+ z'),
\end{eqnarray*}
where $\xi=\frac{1}{2}(1 + z)u + \frac{1}{2}(1+z')(1-u).$

The normalization factor $N_{BSM}$ is determined from the condition $F^{BSM}(0)=1$.
\begin{figure}[h!]
\centering
\includegraphics[width=0.5\textwidth]{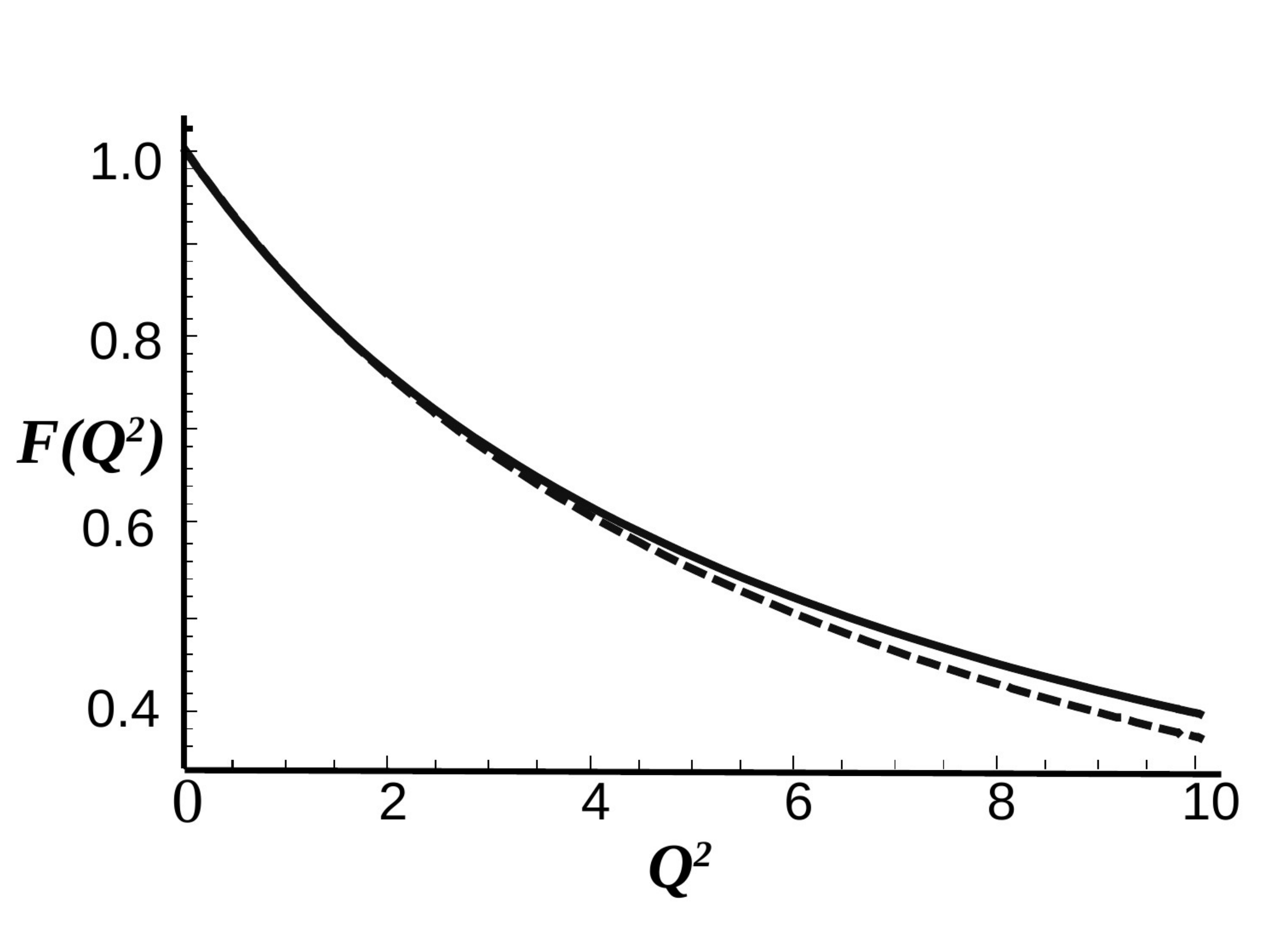}
\caption{EM form factor calculated via LF wave function by Eq.  (\ref{FLFD}), 
for $N_z=2, \;N_x=4$ (dashed curve) and for  $N_z=3,N_x=5$ (solid curve), i.e., via solutions 
$g_E$ shown Fig.  \ref{fg42a}.  }
\label{FFLF}
\end{figure}

The form factor is expressed via LF wave function as follows (see {\it e.g.} Eq. (6.14) from \cite{cdkm}):
\begin{eqnarray}\label{f2b}
F^{LF}(Q^2)&=& \frac{1}{(2\pi)^3} \int \frac{d^2k_{\perp}dx}{2x(1-x)}
\nonumber\\
&\times&\psi_{LF}(\vec{k}_{\perp},x)\psi_{LF}(\vec{k}_{\perp}-x\vec{Q}_{\perp},x) , 
\end{eqnarray} 
where ${\vec Q}_{\perp}^2=Q^2$.
Substituting in (\ref{f2b}) the  LF wave function
$\psi_{LF}(\vec{k}_{\perp},x)$ determined by Eq.~(\ref{lfwf1}), one finds
(see Eq. (26) from \cite{ckm_2009}):
\small
\begin{eqnarray}
&&F^{LF}(Q^2)=\frac{1}{2^5 \pi^3N_{LF}}\int_0^{\infty}d\gamma'
\int_0^{\infty}d\gamma \int_0^1 d x\int_0^1 du
\nonumber\\
&\times &
\frac{x(1-x)\, u(1-u)\,  g(\gamma,2x-1)g(\gamma',2x-1) }
{\left[u\gamma+(1-u)\gamma'+u(1-u)x^2 Q^2+
m^2-x(1-x)M^2\right]^3}.
\nonumber\\
&&
 \label{FLFD}
\end{eqnarray}
\normalsize
The EM form factor calculated via LF wave function by Eq.  (\ref{FLFD}), for $N_x=4, \;N_z=2$ (dashed curve) 
and for  $N_x=5,N_z=3$ (solid curve), i.e., via the solutions for $g_E$ given in 
Fig. \ref{fpsi42}, is shown in Fig.~\ref{FFLF}.  We see that the apparently distinct solutions 
 shown in Fig. \ref{fg42}
 does not result in a considerable difference of the  corresponding form factors. 
Notice that the differences represent only a 5\% deviation at $Q^2\sim 10 \,m^2 $

\begin{figure}[h!]
\centering
\includegraphics[width=0.5\textwidth]{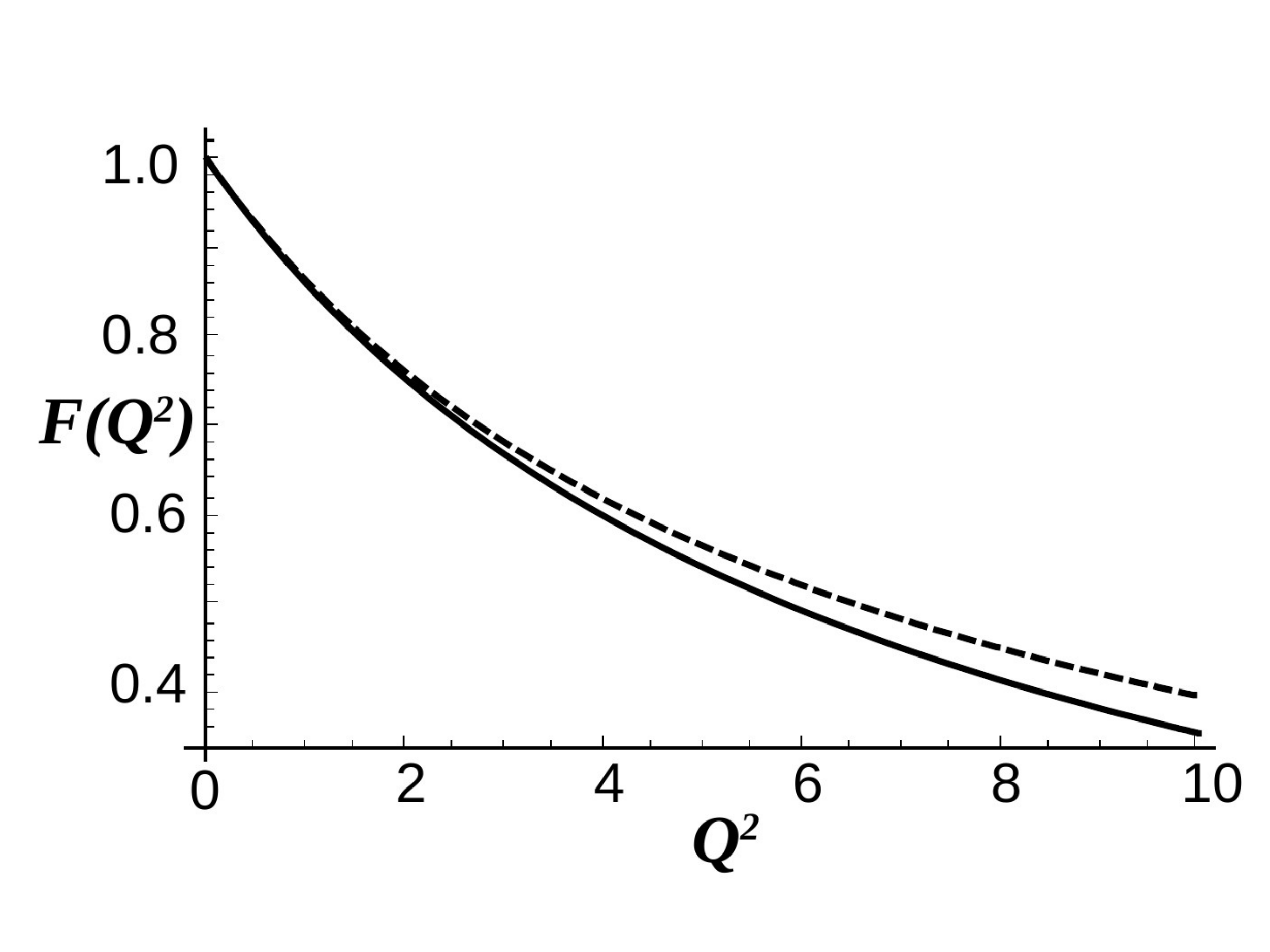}
\caption{EM form factor calculated via Minkowski BS amplitude, by Eq.  (\ref{ffM}) with the same  $g_E$ used to calculate the curves in Fig. \protect{\ref{FFLF}}; the curves are indicated as in Fig. \protect{\ref{FFLF}}.}
\label{FFBS}
\end{figure}

The same form factor, calculated  via the same Nakanishi weight functions, but in the Minkowski BS framework by Eq. (\ref{ffM}) is shown in  Fig. \ref{FFBS}. The difference between the form factors corresponding  to the two solutions 
for $g_E$  shown in  Fig. \ref{fg42} is larger than in Fig. \ref{FFLF}, though it is still not so significant. 

\begin{figure}[t!]
\centering
\includegraphics[width=0.5\textwidth]{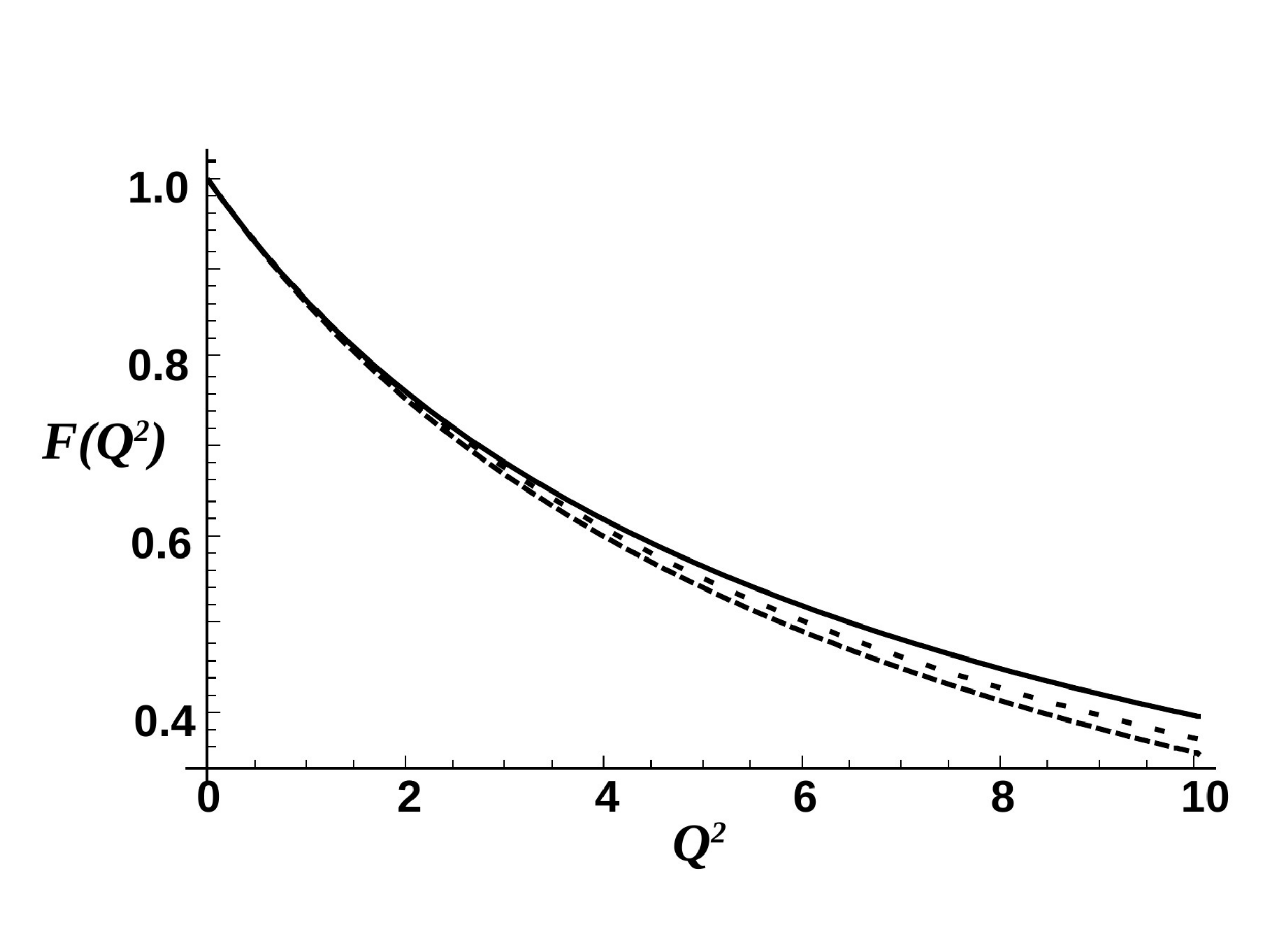}
\vspace{-.5cm}
\caption{The EM form factor for the four calculations given in Figs.
\ref{FFLF}  and \ref{FFBS} with $g_E$ shown in  Fig.  \ref{fg42a}.
{\it Solid (upper) curve:} results obtained  via LF wave function by Eq.
(\ref{FLFD}), for $N_x=5,\;N_z=3$ and via Minkowski BS amplitude,
by Eq.  (\ref{ffM}), for $N_x=4,\;N_z=2$, which are indistinguishable
within thickness of the line.
{\it Dotted (middle) curve:} results obtained  via LF wave function by Eq.
(\ref{FLFD}) for $N_x=4,\;N_z=2$.
{\it Dashed (bottom) curve:} results obtained via Minkowski BS amplitude,
by Eq.  (\ref{ffM}),  for $N_x=5, \;N_z=3$.}\label{FFLFBS}
\end{figure}

All four versions from Figs. \ref{FFLF} and  \ref{FFBS} are shown for comparison together  in Fig. \ref{FFLFBS}. One can distinguish
only three curves of the four since two of then coincide with each other.
We compare form factors calculated via LF wave function and Minkowski BS amplitude using $g_E$. The form factor
calculated via LF wave function by Eq.  (\ref{FLFD}), for $N_x=5, \;N_z=3$   and calculated  via Minkowski BS amplitude, by Eq.  (\ref{ffM}), for $N_x=4, \;N_z=2$ are  shown by the solid line (upper curve). They are indistinguishable within thickness of lines.

It is worth to make the two following remarks. 
({\it i}) Though the form factors $F^{LF}(Q^2)$ and $F^{BSM}(Q^2)$ turned out to be very close to each other (as it is seen 
from the present calculations and from \cite{ckm_2009}), they should not coincide exactly. Their deviation seen in
 Fig. \ref{FFLFBS} is caused not only by numerical uncertainties, but also by physical reasons. This deviation cannot be completely 
 removed by more precise calculations. From point of view of the Fock decomposition of the state vector, the form factor 
$F^{LF}(Q^2)$ is determined  only by the two-body contribution in the state vector, whereas $F^{BSM}(Q^2)$, determined by the 
two-body BS amplitude, includes implicitly the effect of higher LF Fock states. 
({\it ii}) The contribution of the higher Fock components can be found knowing only the two-body component. Indeed, inverting Eq. 
(\ref{lfwf1}), we extract $g_{LF}$ from the two-body LF wave function $\psi_{LF}$. It is the same $g$ that enters in the 
Minkowski BS amplitude (\ref{nak}) and in the form factor  (\ref{ffM}), including the higher sector contributions.

As it was just mentioned,  the form factor  (\ref{ffM}) (calculated with $g_{LF}$ extracted from the two-body LF wave function) 
includes implicitly not only the two-body component contribution but also the higher Fock components. 
So, this higher Fock sector contribution  is found from the two-body component which was taken as input. 
Though it seems a little bit paradoxical (the possibility to get information about higher Fock states from the two-body one), this is, probably, a manifestation of self-consistency of the Fock decomposition embedded in the
Nakanishi integral representation.
In the field-theoretical framework (in which only the BS amplitude can be defined via the Heisenberg field operators \cite{SB_PR84_51}), 
the number of  particles is not conserved and the existence of one Fock component
requires the existence of other ones. The set of them, corresponding to different numbers of particles,  ensures also the 
correct transformation properties of the full state vector $|p\rangle$ since the Fock components are transformed by 
dynamical LF boosts in each other. 

The fact that the difference between form factors determined by the two-body LF wave function and the form factor including higher components is small was found also in Wick-Cutkosky model \cite{dshvk}.  The contribution of many-body components with $n \ge 3$ reaches 36\% in the full normalization integral $F(0)$ only for the coupling constant so huge that the total mass of the bound system tends to zero.

 A final remark: the stability of the form factors  obtained from $g_E$, though
  quite unexpected in view of the sizeable difference of the Nakanishi weight functions (see e.g. Fig. \ref{fpsi42}), is in fact
 appropriate. As we have discussed in the analytical example, the reason for the instability of the solution of the linear system
 is the contribution from the very small eigenvalues. Therefore the difference between the $g_E$'s comes from the
 corresponding eigenvectors  which induce the observed oscillations. 
 However, when using the Minkowski BS amplitude to compute the form factor, 
 the contributions from the eigenvectors with small eigenvalues are damped in the same way they were enhanced in the inversion. 
 It is very likely that this result follows from the fact that the spectra of the
 Nakanishi kernels  in Minkowski and Euclidean spaces are the same.

%%%%%%%%%%%%%%%%%%%%%%%%%%%%%%%%%%%%%%%%%
\section{Discussion and conclusion}\label{concl}

 We have demonstrated by explicit calculations, that the Nakanishi representations  of the Bethe-Salpeter amplitude in 
Euclidean space  ($\Phi_E$) and  the light-front wave function  ($\psi_{LF}$) can be numerically inverted.
If one of this two quantities is known, one can  easily calculate  the other one as well as
 the Bethe-Salpeter amplitude in Minkowski space  ($\Phi_M$)  and associated observables, like the electromagnetic form factor.
 
We have developed an analytically solvable model, in the framework of which we  compared 
the accuracy of the numerical solution with the analytical one. 
Though the  inversion of a Fredholm integral equation 
of the first kind providing $g$ is an ill-posed problem, it  
can be solved with satisfactory precision by using appropriate methods. 

Our best results are obtained with the Tikhonov regularization procedure.
This method is rather efficient and it allowed to successfully overcome the instabilities of the solution which we  found 
in a previous work \cite{FCGK} when computed $g$ from $\Phi_E$.  In addition, it turns out that the light-front
wave function and observables are insensitive to 
the residual uncertainty of $g$ which remains after stabilizing the solution by the regularization. 
The uncertainties on the Light-Front wave function  are within less than 1\%, that is much smaller than for $g's$.

The Nakanishi weight function $g$ was expanded in terms of  Gegenbauer polynomials. 
A small number of terms in this decomposition is required to control the instabilities.
This method is very efficient for describing monotonic behaviors  but it is not sufficient to reproduce more involved structure of $g$ like the ones provided by the dynamical model.  
The later ones generate considerable uncertainties in the inversion procedure.
Increase of the number of terms in the decomposition gives more flexibility, but, at the same time, it enhances
the numerical instability of the solution. We believe that it would be useful to study other 
basis  or discretization methods,  reflecting more the particular functional form of $g$. 

Our approach can find  interesting applications to extract Minkowski amplitudes
from an Euclidean theory, like for instance Lattice QCD.
The Euclidean BS amplitude is being currently computed there with the
full dynamical contents of the theory.
If one is able to  extract from it the corresponding Nakanishi weight function $g$, one can access to
time-like form factors, momentum distributions, GPD's and TMD's which are not accessible in a direct way.  
This could considerably simplify the study of the Minkowski space structure of hadrons from Lattice QCD ab-initio calculations.

%%%%%%%%%%%%%%%%%%%%%%%%%%%%%%%%%%%%%%%
\section*{Acknowledgements}
We thank G. Salm\`e for useful discussions.
One of the authors (V.A.K.) acknowledges 
the grant \#2015/22701-6 from Funda\c c\~ao
de Amparo \`a Pesquisa do Estado de S\~ao Paulo (FAPESP). 
He is also sincerely grateful to group of theoretical nuclear physics of 
ITA, S\~{a}o Jos\'e dos Campos, Brazil, for kind hospitality during his visit.

%%%%%%%%%%%%%%%%%%%%%%%%%%%%%%%%%%%%%%%%%%%%%%%%%%


\begin{thebibliography}{99}

\bibitem{SB_PR84_51}
E.E.~Salpeter, H.A.~Bethe, Phys. Rev. {\bf 84}, 1232 (1951)

\bibitem{Guernot_Eichmann}			
G. Eichmann, Few Body Syst. {\bf 57}, (2016) 541 (2016);  
Phys. Rev. D {\bf 84}, 014014 (2011)

%\highlight{G. Eichmann, Calculating the Euclidean BS amplitude with complex arguments}
%Towards a microscopic understanding of nucleon polarizabilities

\bibitem{KusPRD} K. Kusaka, A. G. Williams, Phys. Rev. D {\bf 51}, 7026 (1995);\\
K. Kusaka, K. Simpson, A.G.~Williams, Phys. Rev. D {\bf 56}, 5071 (1997)
\bibitem{bs1}
V.A.~Karmanov, J.~Carbonell, Eur. Phys. J.  A  {\bf 27}, 1 (2006);\\  J. Carbonell, V.A. Karmanov, 
%{\it Cross-ladder effects in Bethe-Salpeter and Light-Front equations},   
Eur. Phys. J.  A {\bf 27}, 11 (2006)

\bibitem{FrePRD12} 	
T. Frederico, G. Salm\`e, M. Viviani, Phys. Rev. D \textbf{85}, 036009 (2012); Phys. Rev. D {\bf 89}, 016010 (2014)

\bibitem{fsv_2014}
T.~Frederico, G.~Salm\`e, M.~Viviani, Phys. Rev. D {\bf 89}, 016010 (2014)

\bibitem{nakanishi} N.~Nakanishi, Prog. Theor. Phys. Suppl. {\bf 43}, 1 (1969);
{\bf 95}, 1 (1988)

\bibitem{bs-long}
J.~Carbonell, V.A.~Karmanov, 
%{\it Solving Bethe-Salpeter scattering state equation in Minkowski space}, arXiv:1408.3761;  
Phys. Rev. D {\bf 90}, 056002 (2014)% (4 September).

\bibitem{cdkm} 
J. Carbonell, B. Desplanques, V.A. Karmanov, J.-F. Mathiot, Phys. Reports, {\bf 300} (1998) 215

\bibitem{MC_2000} 
M. Mangin-Brinet and J. Carbonell, Phys. Lett. B {\bf 474}, 237 (2000)

 \bibitem{Brodsky98}  
S.J.~Brodsky, H.~Pauli, and S.S.~Pinsky, Phys. Rep. {\bf 301}, 299 (1998)

\bibitem{vary2016}
J. P. Vary, L. Adhikari, G. Chen, Y. Li, P. Maris and X. Zhao, 
%Basis Light-Front Quantization: Recent Progress and Future Prospects
Few Body Syst. {\bf 57}, 695 (2016)

\bibitem{Craig}
Lei Chang, I. C. Clo\"et, J. J. Cobos-Martinez, C. D. Roberts, S. M. Schmidt, and P. C. Tandy,
%Imaging Dynamical Chiral-Symmetry Breaking: Pion Wave Function on the Light Front
Phys. Rev. Lett. {\bf 110}, 132001 (2013)

\bibitem{FCGK}
T. Frederico, J. Carbonell, V. Gigante and V.A.~Karmanov, 
%{\it Inverting the Nakanishi integral relation for a bound state Euclidean Bethe-Salpeter amplitude}, Contribution to the proceedings of the Workshop: Light-Cone %2015, Frascati, Italy, September 21-23, 2015. 
Few-Body Syst. {\bf 56}, 549 (2016)

\bibitem{Efros_SJNP_1985} V.D.  Efros,  Sov. J. Nucl. Phys. {\bf 41}, 949 (1985) 
 
\bibitem{Efros:2007nq} V. D. Efros, W. Leidemann, G. Orlandini, and N. Barnea, J. Phys. G: Nucl. Part. Phys. {\bf 34}, R459 (2007) 

\bibitem{Giussepina_EFB23_2016} G. Orlandini, F. Turro, arXiv:1612.00322   
 %Integral transform methods: a critical review of various kernels

\bibitem{Herve}  C. Mezrag, H. Moutarde, J. Rodriguez-Quintero, Few Body Syst. {\bf 57}, 729 (2016)   
%From Bethe-Salpeter Wave functions to Generalised Parton Distributions


\bibitem{BS_Latt}  N. Ishii, S. Aoki, T. Hatsuda, Phys. Rev. Lett. {\bf 99}, 022001 (2007)

\bibitem{MEM}
Fei Gao, Lei Chang, Yu-xin Liub, 
%{\it A Novel Algorithm for Extracting the Parton Distribution Amplitude from the Euclidean Bethe-Salpeter Wave Function},
arXiv:1611.03560 [nucl-th]

\bibitem{Tikhonov}
A.N.~Tikhonov, A.V.~Goncharsky, V.V.~Stepanov, A.G.~Yagola,  {\it Numerical Methods for the Solution of Ill-Posed Problems}, Kluwer Academic Publishers, 1995
 
\bibitem{Gianni} 
G.~Salm\`e, private communication

%\bibitem{Jaume}
%J.~Carbonell, private communication.

\bibitem{ckm_2009}
J.~Carbonell, V.A.~Karmanov, M.~Mangin-Brinet,
%{\it Electromagnetic form factor via Bethe-Salpeter amplitude in Minkowski space},
%Comments: 8 pages, 7 figures, 
Eur. Phys. J. A {\bf 39}, 53  (2009)  %-60; arXiv:0809.3678 (hep-th).

\bibitem{dshvk}
Dae Sung Hwang and V.A. Karmanov, 
%{\it Many-body Fock sectors in Wick-Cutkosky model}, 
Nucl. Phys. B {\bf 696}, 413 (2004)


\end{thebibliography}
\end{document}